\documentclass[twocolumn,showpacs,superscriptaddress,amsmath,amssymb]{revtex4}
\usepackage{stmaryrd}
\usepackage{mathrsfs}
\usepackage{graphicx}
\usepackage{dcolumn}
\usepackage{bm}
 
\def\bd{\begin{document}} \def\ed{\end{document}}
\def\bmp{\begin{minipage}} \def\emp{\end{minipage}}
\def\bcc{\begin{center}} \def\ecc{\end{center}}     \def\npg{\newpage}
\def\beq{\begin{equation}} \def\eeq{\end{equation}} \def\hph{\hphantom}
\def\be{\begin{equation}} \def\ee{\end{equation}} \def\r#1{$^{[#1]}$}
\def\n{\noindent} \def\ni{\noindent} \def\pa{\parindent}
\def\hs{\hskip} \def\vs{\vskip} \def\hf{\hfill} \def\ej{\vfill\eject}
\def\cl{\centerline} \def\ob{\obeylines}  \def\ls{\leftskip}
\def\underbar#1{$\setbox0=\hbox{#1} \dp0=1.5pt \mathsurround=0pt
   \underline{\box0}$}   \def\ub{\underbar}    \def\ul{\underline}
\def\f{\left} \def\g{\right} \def\e{{\rm e}} \def\o{\over} \def\d{{\rm d}}
\def\vf{\varphi} \def\pl{\partial} \def\cov{{\rm cov}} \def\ch{{\rm ch}}
\def\la{\langle} \def\ra{\rangle} \def\EE{e$^+$e$^-$} \def\pt{p_{\rm t}}
\def\dt{\delta}   \def\ie{{\it i.e.\;}}   \def\cf{{\it cf.\;}}
\def\bitz{\begin{itemize}} \def\eitz{\end{itemize}}
\def\btbl{\begin{tabular}} \def\etbl{\end{tabular}}
\def\btbb{\begin{tabbing}} \def\etbb{\end{tabbing}}
\def\beqar{\begin{eqnarray}} \def\eeqar{\end{eqnarray}}
\def\\{\hfill\break} \def\dit{\item{-}} \def\i{\item}
\def\bbb{\begin{thebibliography}{9} \itemsep=-1mm}
\def\ebb{\end{thebibliography}} \def\bb{\bibitem}
\def\bpic{\begin{picture}(260,240)} \def\epic{\end{picture}}
\def\akgt{\noindent{Acknowledgements}}   \def\om{\omega}
\def\fgn{\noindent{\bf\large\bf figure captions}}
\def\la{\langle}   \def\ra{\rangle}  \def\bt{\beta}
\def\ifmath#1{\relax\ifmmode #1\else $#1$\fi}%
\def\rc{\ifmath{{\mathrm{c}}}}
\def\cut{\ifmath{{\mathrm{cut}}}}
\def\rF{\ifmath{{\mathrm{F}}}}
\def\rK{\ifmath{{\mathrm{K}}}}
\def\rp{\ifmath{{\mathrm{p}}}}
\def\rt{\ifmath{{\mathrm{t}}}}
\def\LAB{\ifmath{{\mathrm{LAB}}}}
\def\cut{\ifmath{{\mathrm{cut}}}}
\def\beq{\begin{equation}}
\def\eeq{\end{equation}}
\def\bcs{\begin{cases}}  \def\ecs{\end{cases}}
\def\us{^{(s)}}  \def\bea{\begin{eqnarray}} \def\eea{\end{eqnarray}}
\def\nbr{\nonumber} \def\e{{\rm e}} \def\dt{\delta} \def\D{\Delta}
\def\r{\rho}  \def\unln{\underline}
\newcommand{\cinst}[2]{$^{\mathrm{#1}}$~#2\par}
\newcommand{\crefi}[1]{$^{\mathrm{#1}}$}
\newcommand{\crefii}[2]{$^{\mathrm{#1,#2}}$}
\newcommand{\crefiii}[3]{$^{\mathrm{#1,#2,#3}}$}
\newcommand{\HRule}{\rule{0.5\linewidth}{0.5mm}}   \def\af{\alpha}
\def\fc{\frac}   \def\sNN{\sqrt{s_{\rm NN}}}   \def\ve{\varepsilon}
\def\yw{Y_{\rm w}}  \def\ew{\eta_{\rm w}}  \def\de{\delta\eta} \def\Bs{B_{\rm s}}
\def\bs{\boldsymbol}  \def\lm{\lambda} \def\pinf{{\rm P}_\infty}
\def\tmx{T_{\rm max}}
\def\ec{\eta_{\rm c}}
\def\tmn{T_{\rm min}}
\def\rcmi{r^{\rm cm}_i}
\def\mb{\mu_B}   \def\tfz{T_{\rm fz}}
\def\tcr{T_{\rm cr}}   \def\mcr{{\mb}_{\rm cr}}
\def\ptflow{\bs p_{\rm t}^{\rm flow}}  \def\ptherm{\bs p_{\rm t}^{\rm therm}}
\def\ptvib{\bs p_{\rm t}^{\rm vib}} \def\sm{\sigma}

\bd

\title{The characterizing variable for critical point in momentum space}

\author{Du Jiaxin}

\affiliation{Institute of Particle Physics, Huazhong Normal
University, Wuhan 430079 China}

\author{Ke Hongwei}

\affiliation{Institute of Particle Physics, Huazhong Normal
University, Wuhan 430079 China}

\author{Xu Mingmei}

\affiliation{Institute of Particle Physics, Huazhong Normal
University, Wuhan 430079 China}

\author{Liu Lianshou}
\email{liuls@iopp.ccnu.edu.cn}

\affiliation{Institute of Particle Physics, Huazhong Normal
University, Wuhan 430079 China}

\affiliation{Key  Laboratory of Quark \& Lepton Physics (Huazhong
Normal University), Ministry of Education, China}

\def\ssnn{\sqrt {s_{NN}}}

\begin{abstract}
The possible experimentally observable signal in momentum space for
the critical point, which is free from the contamination of
statistical fluctuations, is discussed. It is shown that the higher
order scaled moment of transverse momentum can serve as an
appropriate signal for this point, provided the transverse momentum
distribution has a sudden change when energy increases passing
through the critical point. A 2-D percolation model with a linear
temperature gradient is constructed to check this suggestion. A
sudden change of third order scaled moment of transverse momentum is
successfully observed.
\end{abstract}


\maketitle

\section{Introduction}

Heavy-ion collision experiments at the relativistic heavy ion
collider BNL-RHIC have found evidences that a phase transition from
hadron gas to quark gluon plasma (QGP) has occurred~\cite{evidence}.
Theoretical studies predict the existence of a critical point, or
critical end point (CEP)
--- a point which separates the first order phase transition at high
baryon chemical potential, low temperature from the smooth crossover
at low baryon chemical potential, high temperature~\cite{cep}.
Theoretical estimation strongly indicates that, if the critical end
point does exist, it is within the region of the phase diagram
probed by the heavy ion collision experiments. Most probably it will
appear on the quark-gluon phase transition boundary in the range of
baryon chemical potential of 100-500 MeV, which corresponds to heavy
ion collisions at $\sqrt{s_{NN}}$= 5 -- 50 GeV~\cite{estimate}. The
two collaborations in RHIC --- STAR and PHENIX --- have already
begun their procedure of low energy scan to observe the critical
point~\cite{scan}.

In order to locate the critical point we must have some variable as
appropriate signal. This variable should be determined by the
thermodynamic property of critical point. Two distinct properties of
the critical point are:

A,\ {\it Some quantities have large fluctuations at the critical
point}. A well-known example is the critical  opalescence, which
appears at the critical point of liquid-gas system, indicating the
existence of large density fluctuations.

B,\ {\it Some quantities change from zero to non-zero at the
critical point}. For example, in a ferro-magnetic system the global
magnetization changes from zero to non-zero at the critical point.

These two properties are closely related in thermodynamics, but
phenomenologically they have different representations and lead to
different experimental observables.

In the study of relativistic heavy ion collision the above-mentioned
property A has been widely utilized by many authors to propose
signals for critical point\;\cite{signalA}. It is expected that
large fluctuations exist at the critical point for many variables,
such as transverse momentum, K/$\pi$ ratio, etc. What is important
is that these variables have to be measured event by event. Due to
the limited number of particles in a single event, they have the
disadvantage of being inevitably contaminated by the statistical
fluctuations\;\cite{statfluc1, statfluc2, statfluc3}. There is no
reliable method to eliminate the statistical fluctuations and it is
unclear whether any fluctuation signal for critical point can
survive after eliminating the statistical fluctuations.

In the present letter we will discuss the possibility of finding
some signal for the critical point, basing on its property B. Such a
variable, if exists, will be measured in the whole event sample
instead of event by event. Therefore, it is free from the
troublesome problem of statistical fluctuations. The statistical
errors can be reduced by enlarging the statistics, i.e. increasing
the number of events in the sample.

The layout of the paper is as follow: In Sect. 2, we discuss the
moments of final particle momentum distribution, which may be
candidates for such characterizing variable. Then, in Sect. 3, a
2-dimensional site-percolation model with temperature gradient,
 which can be used to exhibit this characterizing variable, is presented and
 discussed.
Finally. The summary and conclusion is presented in Sect. 4.

\section{Moments of final particle momentum distribution}

In trying to find a measure basing on the property B of critical
point it should be noticed that the only quantities that are
observable in relativistic heavy ion experiments are the momentum,
mass, charge, etc. of final state particles. On the contrary the
variables that changes from zero to non-zero at the critical point
is usually in the coordinate space. The aim of this letter is to
find out a variable in momentum space that can characterize the
critical behavior in coordinate space.

Let us ask a question: what information can be extracted from the
transverse momentum of final-state particles. The answer is: all the
available information about transverse momentum is contained in its
probability distribution, or equivalently, in the moments of the
distribution of all orders. Among them are:

First order moment $\la \pt\ra$ gives the average transverse
momentum, where $\la \cdot \ra$ means average over the whole event
sample.

Second order moment $\la \pt^2\ra$ describes the width of the
distribution. Measuring in each event and combining with the first
order moment, it gives the event-by-event $\pt$-fluctuation
$\sm^2=\overline{\pt^2}-{\overline\pt}^2$, where $\overline{\pt^2}$
and ${\overline\pt}$ are the average of $\pt^2$ and $\pt$ in each
event, respectively.

Third order moment $\la\pt^3\ra$ indicates the peak of the
distribution.

Fourth and other higher order moments describe the detailed shape of
the distribution.

\begin{figure}
\includegraphics[width=2.2in]{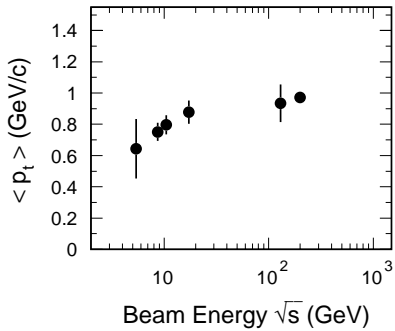}
\caption{\label{figstar}  The average transverse-momentum of $\phi$
in A+A collisions at different colliding energies. Data taken from
Ref.~\citep{STARpT}.}
\end{figure}

As discussed above the event-by-event fluctuation
$\sm^2=\overline{\pt^2}-{\overline\pt}^2$ has been suggested by many
authors as a possible signal for critical point. It has the
disadvantage of being contaminated by statistical fluctuations. So
we turn to the first order and the higher than second order moments.
The present available experimental data on $\la\pt\ra$ of $\phi$
particle show a clear rise and subsequent saturation as the
increasing of colliding energy, as depicted in Fig.\;\ref{figstar}.
This monotonic behavior, if exists also for other identified
particles, e.g. pions, is encouraging. it indicates that possibly an
abrupt jump has occurred at a certain energy.

Let us define the {\it scaled n-th order moment} and its reciprocal
as \beq C_n=\fc{\la \pt^n\ra}{\la \pt\ra^n}, \qquad
D_n=\left(C_n\right)^{-1}= \fc{\la \pt\ra^n}{\la \pt^n\ra}.\eeq Both
of them are dimensionless and are, therefore, more suitable to serve
as a characterizing variable. The higher than second order scaled
moments portray the shape of transverse momentum distribution. If
when the energy increases passing through the critical point there
were a sudden change in shape of the transverse momentum
distribution, then the higher order scaled moments of $\pt$ might
have a visible jump. This phenomenon if exists can serve as a good
candidate of the critical-point signal.

Here comes another advantage of this approach in comparison with the
usual one based on the large fluctuations at the critical point. As
we know, the first round of energy scan will be performed with large
steps between different colliding energies. If it is not by occasion
that some energy used in the first round just locates at the
vicinity of the fluctuation peak, we will see nothing in the first
round and the subsequent scan has to be carried out in finer steps
over the whole energy range. On the contrary, if the higher order
scaled $\pt$ moments have a sudden rise when passing through the
critical point, then already in the first round of energy scan we
will observe a rise of these moments, and most probably the critical
point is located in the region of the moment-rising. Then the
further scan could be concentrated in this region and the critical
point, signaled by the abrupt jump of the scaled moments, can be
catched easily in this way.

\section{2-dimensional site-percolation model with temperature gradient}

Let us use a simple well-defined model that has known critical
behavior to exhibit the above argument. The model is a 2-dimensional
site-percolation model\cite{perc} with temperature gradient. We take
a big circle of radius $R$ and randomly locate in it $n$ little
circles of radius $r$ ($r\ll R$). These little circles will be
referred to as cells. If the number of cells is large enough, two or
more cells may overlap. In this case, a connection is built among
these cells and a cluster is formed. Clusters can be of various
size. If there is a cluster extending from one side of the big
circle to the other side, then it is referred to as an {\it infinite
cluster}. The appearance of an infinite cluster is considered as the
appearance of a new phase.

The size of the system is defined as the radius of the big circle
with that of the small circle as unit: \beq L=\frac{R}{r}.\eeq In
the following calculation, we will fix the size of the system and
assign $L$ the value 500. The model parameter is chosen to be the
sum of the area of all the small circles and the area of the big
circle: \beq \eta=\frac{n\pi r^2}{\pi R^2}=\fc{n}{L^2}. \eeq In the
real physical system the parameter $\eta$ corresponds to the energy
density of the system.

\begin{figure}
\includegraphics[width=2.2in]{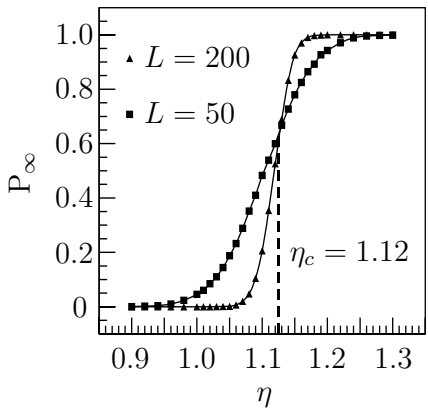}
\caption{\label{pinf} The dependence of $\pinf$ on $\eta$.}
\end{figure}

In Fig.\;{\ref{pinf}} is shown the dependence of the probability
$\pinf$ of the appearance of infinite cluster on the parameter
$\eta$. It can be seen from the figure that, when $\eta$ is small,
infinite cluster almost does not appear, but as $\eta$ increases
when $\eta$ goes near to a certain value $\ec$, the probability
$\pinf$ of infinite cluster increases gradually to unity. The bigger
the system, the sharper the rising of $\pinf$. When $L\to\infty$,
the value of $\pinf$ jumps suddenly from 0 to 1. The value of $\eta$
where this jump appears marks the critical point and is denoted by
$\ec$. In our system, $\ec$ is about $1.12$. It is noticeable that
all the values of $\pinf$ for various system sizes intersect at
$\eta=\ec$. This is a basic property of the percolation model and
can be utilized to locate the critical value $\ec$.

The step-function singularity of $\pinf$\cite{p_inf} is the critical
phenomenon in the model. It is evidently in the coordinate space. In
order to see how this phenomenon represents itself in the momentum
space, we add a temperature gradient into the model, assuming that
the center of the big circle is at a higher temperature $\tmx=400$
MeV, and the fringe gets a lower temperature $\tmn=150$ MeV. A
linear temperature gradient is assumed along the radius of the big
circle.

We further assume that each cluster has arrived at thermal
equilibrium and all the cells in it have the same temperature. The
temperature of the $i$th cluster is determined by the distance to
the center of the big circle of its center of mass defined as \beq
\rcmi=\frac{1}{n_{i}}{\sum_j r_{ij}}, \eeq where $r_{ij}$ is the
distance of the $j$th cell in cluster $i$ to the center of the big
circle, and $n_i$ is the number of cells in cluster $i$, Thus, the
temperature of all the cells in cluster $i$ is: \beq
 T_{i}=\tmx-\frac{\rcmi}{R}\left
(\tmx-\tmn\right ) \eeq

In Fig.\;\ref{t1} are shown the distributions of the number of cell
at different temperatures.

\begin{figure}
\includegraphics[width=3in]{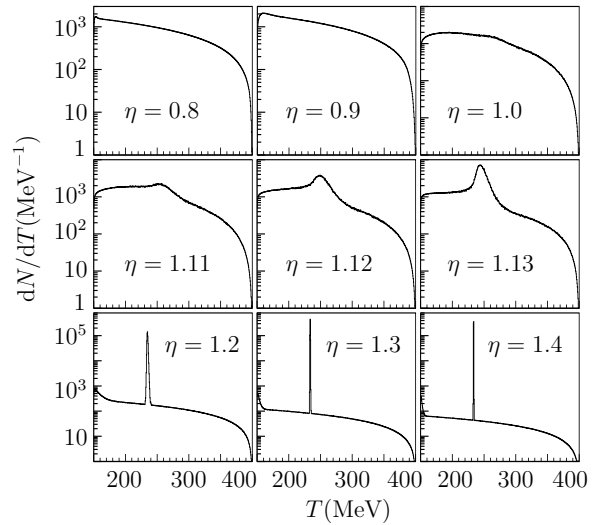}
\caption{\label{t1}The distribution of cell temperature for
different $\eta$.} \vskip 0.5cm
\end{figure}

From the figures we can see that, as the increase of $\eta$, when
$\eta\approx 1.12$ a peak appears suddenly in the distribution
around $T=233$ MeV, and the height of the peak increases very fast.

The transverse momentum of the cells in a cluster can be calculated
from the thermal equilibrium momentum distribution, i.e. the
Boltzmann distribution, \beq \frac{{\rm d} N}{2\pi \pt {\rm d}
\pt}=\frac{N}{2\pi mkT}\exp{\left(\frac{-\pt^2}{2mkT}\right)} \eeq
where $k$ is the Boltzmann constant, and $m$ is the mass of the
cell. In our calculation, we assume the mass of cell to be the mass
of pion, equal to $135$ MeV. Calculating the transverse momentum of
each cell from this distribution, the $\pt$ distributions are
obtained and shown in Fig.\ref{p1}.

\begin{figure}
\includegraphics[width=2.8in]{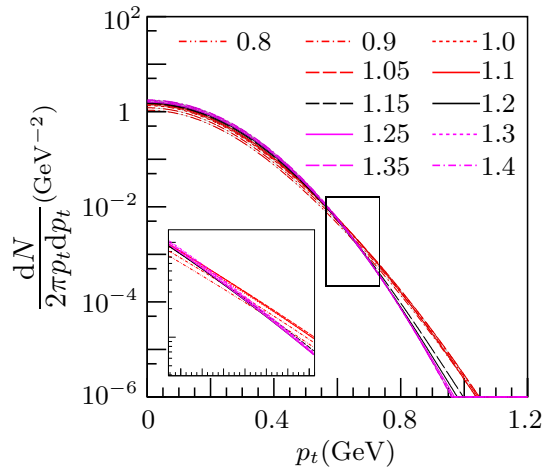}
\caption{\label{p1}(Color on line) The distribution of cell
transverse-momentum for different $\eta$.}
\end{figure}

\begin{figure}
\includegraphics[width=2.2in]{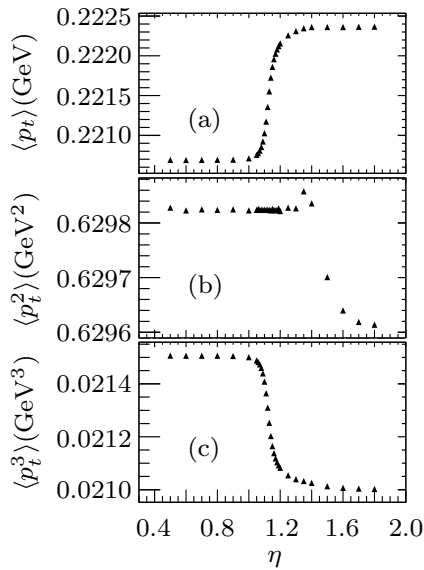}
\caption{\label{sig123}The dependence on $\eta$ of the 1st, 2nd and
3rd order moments of $\pt$ distribution.}
\end{figure}

It can be seen from Fig.\ref{p1} that the $\pt$ distributions for
$\eta<1.15$ and $\eta>1.2$ fall in two groups with a gap in between.
This indicates that there is an abrupt change in the shape of $\pt$
distribution when $\eta$ increases passing through the critical
point.

In order to find a variable that characterizes this abrupt change of
$\pt$ distribution, we calculate the moments of $\pt$ distribution
of first, second and third orders, as shown in Fig.\;\ref{sig123}.
It can be seen from the figures that both the 1st and the 3rd order
moments $\la \pt\ra$ and $\la \pt^3\ra$ change abruptly at the
critical point, while the 2nd order moment $\la \pt^2\ra$ does not.

Although the changes of the first and third order moments
$\la\pt\ra$ and $\la\pt^3\ra$ are in the reverse direction, they
occur at one and the same point, i.e. the critical point. The
dependence of $D_3$  defined in Eq.\;(1) on $\eta$, shown in
Fig.\;6, represents a clear step-function shape. In this figure the
abscissa $\eta$ can be changed to the energy density $\ve$ and the
latter in turn is related to the colliding energy $\sNN$ in heavy
ion collision experiments. We see that a sudden jump of $D_3$ as the
increasing of $\sNN$ can be used as an appropriate signal of
critical point in momentum space.

\begin{figure}
\includegraphics[width=2.2in]{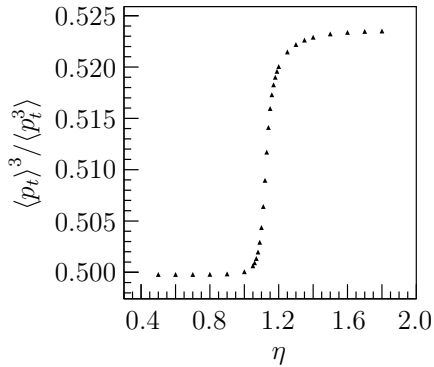}
\caption{\label{Dsig3}The dependence of $D_3$ on $\eta$.}
\end{figure}

As stated above a strict step function singularity exists only in an
infinite system. In the real case of heavy ion collisions the system
is of finite size. The rise of the characteristic variable, $D_3$,
may not be very sharp and may be hard to be recognized. To solve
this problem we can run the experiments with nuclei of various
sizes. The intersection of the experimental results will mark the
place of critical point, cf. Fig.\;2.

\section{Summary and conclusion}

In this letter we try to utilize the sudden change of some
variable(s), instead of the large fluctuations, at the critical
point to find an appropriate variable for locating this point. The
advantage of such an approach is that it is free from the trouble on
the elimination of statistical fluctuations and that the possible
region where the critical point locates can easily be fixed already
in the first rounds of energy scan. Using a site-percolation model
with temperature gradient we found that the third order scaled
moment of $\pt$ distribution (or its reciprocal) can possibly serve
for this purpose.

The model used in this letter is far from a realistic model of
relativistic heavy ion collisions, but the final result is largely
model-independent. Checking the whole derivation carefully we can
see that the essential point for $D_3$ to serve as an appropriate
signal for critical point is the fact that the shape of $\pt$
distribution changes abruptly at the critical point. The higher
order scaled moments (or their reciprocal) represent the shape of
$\pt$ distribution and, therefore, can serve as an appropriate
signal. An abrupt change in $\pt$-distribution shape when
temperature increases passing through the critical point is probable
for the system produced in relativistic heavy ion collisions. The
higher order scaled moment or its reciprocal is, therefore, a
possible candidate for the critical-point signal in these
collisions.

{\bf Acknowledgement} \ Supported by NSFC under projects No.10835005
and 10775056, 10835005.

\ed
\begin{thebibliography}{99}
\bibitem{p_inf}
A real singularity appears only for an infinite system. For a
finite-size system it is smoothed to a rapid rise, cf. Fig.\;2.
\end{thebibliography}
